\def \cm{~\rm{cm}}
\def \s{~\rm{s}}
\def \km{~\rm{km}}

\def \K{~\rm{K}}
\def \g{~\rm{g}}

\def \AU{~\rm{AU}}

\def \yr{~\rm{yr}}

\def \astrobj#1{#1}
\documentclass[12pt,preprint]{aastex}

\shortauthors{Soker}

\begin{document}

\title{A PHENOMENOLOGICAL MODEL FOR THE EXTENDED ZONE ABOVE AGB STARS}

\author{Noam Soker\altaffilmark{1}}

\altaffiltext{1}{Dept. of Physics, Technion, Haifa 32000, Israel;
soker@physics.technion.ac.il.}

\begin{abstract}

I suggest the existence of an extended zone above the surface of
asymptotic giant branch (AGB), as well as similar stars experiencing high
mass loss rates.
In addition to the escaping wind, in this zone there are parcels
of gas that do not reach the escape velocity.
These parcels of dense gas rise slowly and then fall back.
The wind and bound gas exist simultaneously to distances of $\sim 100 \AU$.
I term this region the \emph{effervescent zone}.
In this phenomenological study I find that the density of the bound material
in the effervescent zone falls as $\sim r^{-5/2}$, not much faster than the wind
density.
The main motivation to propose the effervescent model is to allow wide binary
companions to influence the morphology of the descendant planetary nebulae (PN) by
accreting mass from the effervescent zone. Accretion from the effervescent zone is
more efficient than accretion from the wind in forming an accretion disk around
the companion.
The companion might then blow two jets that will shape the descendant PN.
\end{abstract}

\section{INTRODUCTION}
\label{sec:intro}

The transition of asymptotic giant branch (AGB) stars to the early planetary nebula (PN)
stage can last $\sim 100-10^4 \yr$, depending on the mass of the progenitor
(e.g., Sch\"onberner 1983; Bl\"ocker 1995; Frankowski 2003), and on binary interaction
(De Marco 2008).
Binary can have several effects on the final AGB and post-AGB evolution.
A close enough binary companion can enhance the mass loss rate via tidal forces
or a common envelope evolution, both will shorten substantially the post-AGB phase.
A binary companion can lead also to the ejection of mass at a very high rate, part of which
is ejected at speeds below the escape speed; this mass might fall back and extend the
post-AGB phase (Soker 2001b).

It is inferred from the structures of PNs that the last AGB phase involves the transition
from a large scale spherical mass loss geometry to axisymmetrical geometry.
This is generally attributed to interaction with a binary companion (for a recent collection of papers
see Corradi et al. 2008).
Not only close binary companions, but wide companions as well can lead to the formation
of axisymmetrical PNs (Soker 2001a). These wide, orbital separation of $\sim 20-200 \AU$
(depending on eccentricity; Soker 2001a), binary companions do not influence directly the
stelar envelope.
They accrete mass from the AGB wind, form an accretion disk, and blow two opposite jets
that shape the AGB wind to form an axisymmetrical PN at later stages.
For that to occur the AGB mass loss rate should by relatively hight and not
too fast.
The conditions for forming an accretion disk are very sensitive to the outflow speed,
becoming much more favorable for slower outflows (Soker 2001a).
Motivated by this sensitivity I suggest a phenomenological model for an extended
zone above the AGB photosphere where slowly moving mass elements enable
a companion to accrete mass, forms an accretion disk, and blow two jets.
This region is termed the effervescent zone.

The ejection of a slow outflow from AGB stars was discussed before (Soker 2000).
Cool spots in the photosphere of AGB stars lead to enhanced dust formation and
higher mass ejection rate above the cool spot (Soker \& Clayton 1999).
Cool spots are assumed to form by the magnetic activity of the AGB star
(Soker 1998).
The stellar radiation is unable to accelerate the denser gas to the wind speed,
and a region of slower outflowing gas is formed (Soker 2000). In some cases part
of this mass might fall back.
In a previous paper (Soker 2001b) I proposed that mass that was ejected during the final
stage of the AGB falls back to the star while the star is in the post-AGB phase.
For that, the ejected mass reaches distances of hundreds of AU, such that the fall back time is
thousands of years.
For a significant backflow to occur a slow dense flow should exist.
The requirement for both a high mass-loss rate per unit solid angle and a very slow wind,
such that it can be decelerated and flow back, probably requires close binary interaction
(Soker 2001b).
In the present paper I consider ejected mass that reaches a distance of tens of AU rather than
hundreds of AU, such that it falls back before the post-AGB phase.
In the present paper I suggest that several processes that were discussed in previous papers,
and are summarized in section 2, can form an extended region, up to tens of AU, above
the AGB star where parcels of gas rise slowly, and then fall back.
This region might contain also hot, $\sim 10^6 \K$, gas formed by magnetic flares.
The phenomenological model is termed the effervescent wind model.
The density profile in the effervescent zone is derived in section 3.

An extended region above the star where inflow and outflow coexist in the acceleration zone
of the wind has been suggested for massive stars close to their Eddington limit luminosity
(Owocki, van Marle 2008; van Marle et al. 2007).

Single AGB stars can have strong enough mass loss rate to have an effervescent zone
according to the proposed model.
A planet can help in establishing such a zone, as the AGB star can be moderately
spun-up by a planet (Soker 2004), that might lead to
some magnetic activity and enhanced mass loss rate.
A stellar companion at tens to more than a hundred AU can then accrete mass and blow two jets.
A number of such systems, with close planets and a stellar companion at $\sim 20-200 \AU$ are
known (Mugrauer et al. 2007; Chauvin et al. 2007; Correia et al. 2008).

Another motivation for the present study is the \astrobj{Mira AB} symbiotic system.
This is a well studied binary system (e.g., Marengo et al. 2001;
Karovska et al. 1997, 2005; Matthews \& Karovska  2006; Karovska 2006; Wood \& Karovska 2006).
\astrobj{Mira A} is a pulsating AGB star with a radius of $R_1 \simeq 500 R_\odot$ (Wood \& Karovska 2006)
and a mass loss rate and velocity of $\dot M_w \simeq {\rm few} \times 10^{-7} M_\odot \yr^{-1}$
(Ryde \& Sch\"oier 2001)
and $v_w \simeq 7 \km \s^{-1}$ (Knapp et al. 1998), respectively.
There is an asymmetrical circumbinary matter extending to large distances
around the system  (e.g., Planesas et al. 1990).
The region up to tens of AU above the photosphere is quite complicated, with
turbulence (e.g., Ryde \& Sch\"oier 2001), and possibly dense clumps (e.g., Lopez et al. 1997).
The nature of \astrobj{Mira B} is not clear. Either it is a WD or a main sequence star.
Most recently Ireland et al. (2007) argue that it is a main sequence star with a mass of
$\sim 0.7 M_\odot$.
The projected orbital separation is $a \simeq 70 \AU$, but the physical orbital separation
is likely to be $\sim 90 AU$ (Ireland et al. 2007).
Despite the large orbital separation of $a \simeq 40 R_1$ the companion influences the wind
morphology in the region between the two stars
(Marengo et al. 2001; Karovska et al. 1997, 2005; Matthews \& Karovska  2006;
Karovska 2006).
This cannot be attribute to a tidal effect on the photosphere of Mira A,
but rather to an influence on the wind's acceleration zone
(e.g., Mohamed \& Podsiadlowski 2008), or influence on bound mass as suggested here
in section 4.
Section 5 contains a short summary.

\section{THE EFFERVESCENT WIND MODEL}
\label{sec:conditions}

The large scale mass loss geometry during most of the AGB is spherical.
This holds even when there are temporal variations in the mass loss rate.
This is evident from the spherical multiple semi-periodic concentric arcs (rings; shells)
observed around AGB stars (Mauron \& Huggins 2000), in PNs (Corradi et al. 2004),
and in pre-PNs (e.g., Hrivnak et al. 2001).
These shells are at large distances from the center, and show the large scale geometry.

However, close to the star this is not necessarily the case.
Studies of SiO maser clumps above the surface of Mira stars show complicated structures
(e.g. Cotton et al. 2006), and a chaotic motion, with inflow and outflow, that is imposed
on the large scale outflow (Diamond \& Kemball 2003).
Water maser studies, that explore regions further out around AGB stars at $\sim 100 \AU$
(much closer than the distances of the arcs mentioned above), show that the small
scale structure is not homogenous at these somewhat larger distances from the star
(e.g., Vlemmings et al. 2002).
In the present study I am interested in small scale inhomogeneities up to $\sim 100 \AU$
from the star.
The observations listed above and in section 1 concerning Mira A suggest a complicated
structure up to this distance from strongly pulsating AGB stars.

A strong support to an inhomogeneous circumstellar envelope is the
carbon AGB star IRC+10216, which has a complicate zone
extending to tens of AU above the star  (Fonfria et al. 2008 and references therein).
Fonfria et al. (2008) suggest that clumps composed of hot gas are moving inward
and outward at high velocities along different radial directions in the inner
$\sim 10 \AU$ of the circumstellar region. The faster moving clumps then, will reach larger
distances.

Inhomogeneities are observed also in the wind and circumstellar envelope of red supergiant stars
(e.g. Lobel \& Dupree 2000; Humphreys et al. 2007),
which although more luminous, have similar mass loss process to that of AGB stars.
The inhomogeneities include velocities that are different from a radial outflow.
These authors attribute the inhomogeneities to large convective elements (cells; blobs),
or magnetic fields (Humphreys et al. 2007) on the stellar surface.
Observations (e.g., Josselin \& Plez 2007) of granular structure and theoretical studies
(e.g., Schwarzschild 1975; Freytag et al. 2002) show indeed that red supergiants
have a small number of convective elements (blobs) near their surface.
As a result of these large convective elements there are large regions on the surface moving
in and out at about the local sound speed. These will create the inhomogeneities.
The structure of the envelope of stars about to leave the AGB is likely to cause very violent
turbulence (Soker \& Harpaz 1999), which is likely to increase the inhomogeneities in the
outflowing gas.

Another source of inhomogeneities is cool spots on the AGB stellar surface.
Such cool spots might be formed from the convective motion or from magnetic spots
(Soker 1998, 2000).
Enhanced dust formation above the cooler region might lead to higher mass loss
rate from these regions (Soker 1998, 2000; Soker \& Clayton 1999).
When mass loss rate is very high, the stellar radiation will not be able to accelerate
the gas to the escape speed.

The zone above the photosphere might also contain hot $\sim 10^6 \K$ gas, as observed in Mira
(Karovska et al. 2005), and which is thought to result from magnetic flares
(Soker \& Kastner 2003).

Another interesting case is the F5 post-AGB star HD56126.
This post-AGB star has an escape velocity of $\sim 60 \km \s^{-1}$ (Li 2003), and
has simultaneously both inflowing and outflowing gas around it.
The inflow and outflow velocities are up to $\sim 20 \km \s^{-1}$ (Klochkova \& Chentsov 2008),
a non-negligible fraction of the escape speed. This shows that even during the post-AGB phase
stars can have complicated wind structure around them.

The observations and theoretical results described above point to an inhomogeneous zone above
the photosphere of AGB stars, where there are denser and rarefied regions, some of which
escape the star and some that fall back to the star.
I term this zone that lays $\sim 10-100 \AU$ above the photosphere of AGB stars,
{\it effervescent zone}, and the model that describe this region the \emph{effervescent wind model}.
The only postulate is that the zone is inhomogeneous and contains many clumps that will
not escape the star, but rather will fall back.
As demonstrated in section 3 this postulate is sufficient to derive an approximate
general density profile.

The effervescent zone will be important for the evolution when a large fraction of the
mass ejected from the stellar surface is unable to reach the escape speed, i.e., the stellar
radiation cannot accelerate it to escape the star.
This requires a high mass loss rate and a chaotic motion on the photosphere, e.g.,
as that expected from large convective elements and magnetic activity.
The inhomogeneity in the mass loss process can be enhanced by a binary companion,
via tidal forces and by spinning-up the AGB star.
Single AGB stars can also become very chaotic when their envelope mass is low,
as the structure become less stable (Soker \& Harpaz 1999, 2002).
This is mainly because the entropy profile becomes steeper and the density profile
becomes shallower at the end of the AGB, and because the thermal (Kelvin-Helmholtz) time
scale decreases and becomes comparable to the dynamical (pulsation) time scale
(Soker 2008).
The behavior of the photospheric opacity can also contribute to the irregular behavior
(Soker 2006).

The effervescent zone will become extended when the radiation momentum flux is about equal
to the wind momentum flux. At that stage the mass loss rate in the wind is
$\dot M_{wc} = L/(c v_w)$, where $v_w$ is the terminal wind speed and $L$ the stellar luminosity.
Substituting typical values gives
\begin{equation}
\dot M_{wc} = 10^{-5}
\left( \frac{L}{5000 L_\odot} \right)
\left( \frac{v_w}{10 \km \s^{-1}} \right)^{-1}
M_\odot \yr^{-1} .
\label{mwc1}
\end{equation}

In the next section the average (in a spherical shell) density of the bound gas
is found to vary as $r^{-5/2}$, where $r$ is the radial distance measured from the
center of the star.
Namely, the ratio of the average (in a spherical shell) density of the mass in the
effervescent zone that does not escape from the star to the wind averaged density
$\eta(r)$, is found to vary as $\eta(r) \sim r^{-1/2}$.
The average optical depth from the observer down to radius $r$ is
\begin{eqnarray}
\bar \tau = \int_r^\infty  \kappa \rho dr =    3  \  
\left( \frac{v_w}{10 \km \s^{-1}} \right)^{-1}
\left( \frac{\dot M_w}{10^{-5} M_\odot \yr^{-1}} \right)^{-1}
\left( \frac{r}{10 \AU} \right)^{-1}
\left( \frac{\kappa}{10 \cm^2 \g^{-1}} \right)
\nonumber \\
\times
\left[1+\frac{2}{3} \eta(r) \right],
\label{tau1}
\end{eqnarray}
where $\kappa$ is the opacity of the dusty wind.
The effervescent zone becomes extended to tens of $\AU$
when the mass loss rate is high (eq. \ref{mwc1}). From equation (\ref{tau1})
it is seen that the optical depth
to the effervescent zone becomes very large, and its inner parts cannot be observed directly.
For example,  we can take $\eta (r) \sim 1$ at $r \sim 50 \AU$, and find $\bar \tau \sim 1$.
In general, only the outer regions that are dominated by the wind can be observed directly
(depending on the wavelength).
The carbon AGB star IRC+10216, for example,
has such a mass loss rate and cannot be observed directly  (e.g., Fonfria et al. 2008).

Strong pulsations can eject dense clumps to large distances before they are accelerated to
high speeds. For the clumps to fall back the gravitational force must be larger than
that due to radiation pressure. Let us consider a dense clump (a blob), e.g.,
as the one modelled by Lopez et al. (1997) in Mira A.
It has a cross section facing the star of area $A_b$, a length along the
radial direction of $l_b$, ant its density is $k_b$ larger than the ambient
wind density $\rho_w=\dot M_w/4 \pi r^2 v_w$.
The gravitational force on the blob is
\begin{equation}
F_g = \frac{G M_\ast}{r^2}
\frac { k_b A_b l_b\dot M_w}{4 \pi r^2 v_w} ,
\label{fgrav}
\end{equation}
where $M_\ast$ is the mass of the central star.
I assume that the blob is dense and large enough such that most of the radiation is
absorbed by the dusty blob, such that the force due to radiation pressure is
\begin{equation}
F_{\rm rad} = A_b \frac{L}{4 \pi r^2 c}
\label{frad}
\end{equation}
The condition for the blob to fall back is $F_g > F_{\rm rad}$.
Scaling the expression for the forces gives the fall back condition in the form
\begin{eqnarray}
1<\frac{F_g}{F_{\rm rad}}
\simeq
\left( \frac{k_b}{10} \right)
\left( \frac{l_b}{0.5 r} \right)
\left( \frac{r}{50 \AU} \right)^{-1}
\left( \frac{v_w}{10 \km \s^{-1}} \right)^{-1}
\left( \frac{\dot M_w}{10^{-5} M_\odot \yr^{-1}} \right)
\nonumber \\
\times
\left( \frac{\dot M_\ast}{1 M_\odot} \right)
\left( \frac{L}{5000 L_\odot} \right)^{-1} .
\label{ratio1}
\end{eqnarray}

Models that assume clumps exist in the literature.
Lopez et al. (1997), for example, considered a model where dusty clumps denser by a
factor of $\sim 100$ than their environment exist at $50 \AU$ from Mira A.
They didn't consider the dynamical evolution of the clumps, but rather assume
their existence to explain the IR emission of Mira A.
Mira A has a mass loss rate of $\dot M_w \simeq {\rm few} \times 10^{-7} M_\odot \yr^{-1}$
(Ryde \& Sch\"oier 2001), much lower than the scaling value used here.
Still, the extended region above Mira A has clumps and it is turbulence
(e.g., Ryde \& Sch\"oier 2001), as discussed in section 1.
Therefore, it is quite possible that the effervescent model proposed here applies also
to strongly pulsating AGB stars even when their mass loss rate is
$ \dot M_w \ll  10^{-5} M_\odot \yr^{-1}$. For example, if $v_w \sim 5 \km \s^{-1}$
and $k_B \sim 100$ in Mira A.

Dense clumps have larger pressure than their environment, and they expand.
Therefore, very dense clump are losing mass and become smaller.
When the wind mass loss rate is very high and the wind is slow, then for
the blob to fall back it is sufficient for it
to be only slightly denser than the environment.
In any case, the effervescent zone will be extended to a radius of
$R_{e} \sim {\rm few} \times R({\tau=1})$, where $R({\tau=1})$ is the radius where the
optical depth is $\tau=1$ according to equation $\ref{tau1}$.
In cases of strong magnetic activity, e.g., as proposed for Mira A, the effervescent zone
might exist even for optically thin wind, because magnetic activity, e.g., flares and
cool magnetic spots, might also eject material from the AGB star.

\section{THE DENSITY PROFILE OF THE BOUND GAS}
\label{sec:density}

I consider mass ejected from a radius $R_0 \simeq {\rm few} \times R_\ast$ by the
AGB star, where $R_\ast$ is the AGB stellar radius, but stays gravitationally bound to the AGB star.
I assume that the mass ejection rate in an ejection velocity range $dv$ is given by
\begin{equation}
d \dot M_e = f(\tau) \dot M_w \left( \frac{v}{v_{\rm esc}} \right)^{-k_v} \frac{dv}{v_{\rm esc}},
\label{dmdot1}
\end{equation}
where $v_{\rm esc}$ is the escape speed from $R_0$, $\dot M_w$ is the wind mass loss rate
(the mass that has $v>v_{\rm esc}$ and exists simultaneously with the bound mass),
and $k_v$ is a constant of the model.
The effect studied here becomes more significant with increasing optical depth of the wind,
and it is given by an unprescribed factor $f(\tau)$.
It is also assumed that after the ejection no force beside gravity acts on the parcel of gas.
The parcel of gas will reach a radius of
\begin{equation}
R_m=R_o \left(1-\frac{v^2}{v_{\rm esc}^2} \right)^{-1}.
\label{rm1}
\end{equation}
This gives
\begin{equation}
\frac{dv}{v_{\rm esc}} = \frac{1}{2} \left( 1- \frac {R_0}{R_m} \right)^{-1/2}
\frac{R_0}{R_m} \frac{dR_m}{R_m} .
\label{rm2}
\end{equation}
Substituting equations (\ref{rm1}) and (\ref{rm2}) in equation (\ref{dmdot1}) gives
\begin{equation}
d \dot M_e (R_m) = f(\tau) \dot M_w
\left( 1- \frac {R_0}{R_m} \right)^{-(1+k_v)/2} \frac{R_0}{2 R_m} \frac{dR_m}{R_m} .
\label{dmdot2}
\end{equation}
A parcel of mass moving to maximum radius $R_m$ and falling back does it
on a time
\begin{equation}
\Delta t (R_m) = 2^{-3/2} \left(\frac {R_m}{\AU} \right)^{3/2}
\left(\frac {M_\ast}{M_\odot} \right)^{-1/2} \yr.
\label{deltat1}
\end{equation}
The gas spends most of its time near $R_m$.
For the accuracy of this study we can take the parcel of mass to spend all
the time at $R_m$.
Accurate integration that takes the motion of parcels of gas into account will give a density
higher by a factor of $3 \pi/2$, because parcel of gas moving to
distances $r>R_m$ contribute to the density in $R_m$.
This is of no interest here, as the factor is absorbed in $f(\tau)$, and it is better
to bring the simpler and transparent treatment given here.

The mass in a shell of width $d r$ near $R_m$ is then
$d M_e=d \dot M_e(R_m) \Delta t (R_m) $.
Using equations (\ref{dmdot2}) and (\ref{deltat1}) gives the ejected (from the stellar surface)
but bound mass per radial distance $d r$ ($R_M$ for specific parcel of gas is replaced by the
general radial coordinate $r$)
\begin{equation}
\frac {d M_e}{dr} \simeq 2 \times 10^{-6} f(\tau) \frac{R_0}{AU}
\left( \frac{r}{\AU} \right)^{-1/2}
\left(\frac {M_\ast}{M_\odot} \right)^{-1/2}
\left( 1- \frac {R_0}{r} \right)^{-(1+k_v)/2}
\left(\frac {\dot M_w}{10^{-5} M_\odot \yr^{-1}} \right)  M_\odot \AU^{-1}.
\label{dm1}
\end{equation}
I am interested in regions far from the star $r \gg R_0$, such that
$\left( 1- \frac {R_0}{R_m} \right)^{-(1+k_v)/2} \simeq 1$, and I will
omit this term. This gives that ${d M_e}/{dr} \propto r^{-1/2}$ and the average
(over the shell) density $\rho_e (r) \propto r^{-5/2}$.

The mass of the wind in the same spherical shell is
\begin{equation}
\frac {d M_w}{dr} = \frac{\dot M_w}{v_w} =   5 \times 10^{-6} 
\left( \frac{v_w}{10 \km \s^{-1}} \right)^{-1}
\left(\frac {\dot M_w}{10^{-5} M_\odot \yr^{-1}} \right)  M_\odot \AU^{-1}.
\label{dmw}
\end{equation}
The ratio of the bound mass to the wind (escaping) mass
in each spherical shell is equal to the density ratio
\begin{equation}
\eta (r) \equiv \frac{{d M_e}/{dr}}{{d M_w}/{dr}} \simeq 0.4
f(\tau) \frac{R_0}{AU}
\left( \frac{r}{\AU} \right)^{-1/2}
\left(\frac {M_\ast}{M_\odot} \right)^{-1/2}
\left( \frac{v_w}{10 \km \s^{-1}} \right)
\simeq \eta(R_0) \left( \frac{r}{R_0} \right)^{-1/2} .
\label{eta1}
\end{equation}

We can summarize this section as follows.
The fundamental assumption of the model is that mass is ejected at velocities below
the escape speed from a radius $R_0 \simeq {\rm few} \times R_\ast$, and that no other forces
act on the gas beside gravity after ejection.
The main result of the phenomenological model is the finding that an extended
circumstellar zone with an average (over a shell) density profile of
$\rho_e(r) \simeq (r/R_0)^{-5/2}$ is formed.
The gas in this extended zone is rising and falling, but most of the time it is at a very
slow motion.
I term this zone the effervescence zone.
The ratio of the bound gas mass per radial distance to that of the wind is given
by equation (\ref{eta1}).
In general the second equality is not accurate because close to $R_0$ the term
$\left( 1- \frac {R_0}{R_m} \right)^{-(1+k_v)/2}$ is not negligible. It is accurate only
for $k_v = -1$; this implies that more bound mass is ejected at high velocities.
Nonetheless, I take the result of $\eta(r)  \simeq \eta(R_0) ( {r}/{R_0}) ^{-1/2}$,
or $\rho_e \simeq (r/R_0)^{-2.5}$, to
be a general result of the effervescent model.

\section{IMPLICATIONS FOR BINARY INTERACTION}
\label{sec:binary}

Under the assumptions of the phenomenological model the density of the bound material in the
effervescent zone decreases as $\sim r^{-2.5}$.
The effervescent zone will extend to a distance of
$R_{e} \sim {\rm few} \times R({\tau=1})$, where $R({\tau=1})$ is the radius where the
optical depth is $\tau=1$ according to equation $\ref{tau1}$.
In cases of magnetically active AGB stars the zone will be extend even when the optical depth is less
than unity close to the star.
These properties will have to be examine when the physics of wind from AGB stars, e.g., the
coupling of pulsation and dust formation, is better understood, e.g., extending detailed one-dimensional
calculations to three dimensional ones (Sandin 2008).
Here I limit myself to discuss possible implications for the shaping of the circumstellar matter
by a binary companion.
This is after all my main motivation for building the phenomenological effervescent model.

There are two basic modes of mass accretion by a companion outside the AGB envelope:
accretion from the wind (the Bondi-Hoyle-Lyttleton type accretion flow)
and a Roche lobe overflow (RLOF).
Basically, the matter in the wind is already gravitational unbound to the mass losing star
(i.e., it has a velocity above the escape velocity from the mass losing star)
when the companion captures the mass.
In the RLOF-type accretion, on the other hand,
the matter does not reach the escape velocity, and it is captured by the companion
while still being gravitationally bound to the mass losing star.
The RLOF-type accretion is a much more efficient mass transfer process than wind accretion .
The real situation is much more complicated than these two limits of wind accretion and RLOF,
as demonstrated by numerical simulations (e.g.,
Theuns \& Jorissen 1993; Mastrodemos \& Morris 1998,1999; Mohamed \& Podsiadlowski 2008).

A companion at an orbital separation of $a \ga 50 \AU$ cannot accrete via the RLOF from
the AGB star, unless the wind is very slow.
The commonly assumed wind accretion mode is the Bondi-Hoyle-Lyttleton. Only if the orbital
separation is smaller and/or the wind speed is extremely low, a RLOF-type
can occur (Mohamed \& Podsiadlowski 2008).
I use the term RLOF-type because at such large orbital separations synchronization between the
mass losing stellar spin and the orbital motion is not expected (unlike the classical RLOF process).
If a substantial effervescent zone does exist, as is proposed in this paper, then the
companion can accrete from the bound material in the effervescent zone via
a RLOF-type process, simultaneously with the existence of a wind.

Not only the mass accretion rate is relatively more efficient in RLOF-type accretion,
but the accreted mass possesses much higher specific angular momentum.
This material will form an accretion disk around the accreting companion.
The main result of the effervescent zone as far as shaping of the circumstellar matter
is concerned is the formation of an accretion disk. This is likely to result in
launching of two jets that will shape the nebula.

Let us assume that material in the effervescent zone with a maximum radius in the range
$k_1 a<R_m <k_2 a$ is accreted by the companion in a RLOF-type flow, where
$k_1<1$ and $k_2>1$ are two ratios.
The rate of mass injection into this shell is given by integrating the right hand side
of equation (\ref{dmdot2}) from $R_m=k_1 a$ to $R_m =k_2 a$. This gives for $k_v \ne 1$
\begin{equation}
\Delta \dot M_e (k_1 a, k_2 a) = f(\tau) \dot M_w \frac{1}{1-k_v}
\left[ \left( 1- \frac {R_0}{k_2a} \right)^{(1-k_v)/2}
- \left( 1- \frac {R_0}{k_1 a} \right)^{(1-k_v)/2} \right] .
\label{deltame}
\end{equation}
By taking $k_1a=R_0$ and $k_2 a \gg a $ in equation (\ref{deltame})
we get the total injection rate of bound mass into the effervescent zone $\dot M_e$.
Using the inequality $k_1a \gg R_0$, we get then
\begin{equation}
\Delta \dot M_e (k_1 a, k_2 a) \simeq f(\tau) \dot M_w
\frac {R_0}{2a}
\left( \frac {1}{k_1}- \frac {1}{k_2} \right) =
\dot M_e (1-k_v)
\frac {R_0}{2a}
\left( \frac {1}{k_1}- \frac {1}{k_2} \right) .
\label{deltame2}
\end{equation}

As representative numbers let us take $k_1=2/3$, $k_2=2$, $k_v=0$, $a =20 R_0$
(say $R_0=5 \AU$ and $a=100 \AU$), and let the companion
accrete a fraction $\beta\sim 0.5$ of the mass in the shell $k_1 a<R_m <k_2 a$.
The accretion rate from the effervescent zone is then
$\dot M_{\rm acc} = \beta \dot M_e {R_0}/{2a}=0.025 \beta M_e$.
I take the accreting companion to be a main sequence star, that ejects two
jets with a total mass loss rate of $\dot M_j=0.1 \dot M_{\rm acc} $,
and a speed of $v_j=600 \km \s^{-1}$.
The ratio of mass loss rate into the two jets and slow wind for these parameters is
\begin{equation}
\frac{\dot M_j}{\dot M_w} = 5 \times 10^{-4}
\frac {\beta}{0.2}
\frac {\dot M_e}{\dot M_w}.
\label{mjet}
\end{equation}
This ratio is enough for the jets to substantially shape the nebula into a bipolar PN
(Akashi 2008). An accreting WD companion has much faster jets, and can have a much
lower accretion rate and still shape the nebula.
If the accreting main sequence stars accretes at a much lower rate, it might lead to
the formation of an elliptical PN, rather than a bipolar PN (Soker 2001a).

\section{SUMMARY}
\label{sec:sum}

This paper deals with a phenomenological study of a possible extended region
of bound gas around AGB stars experiencing high mass loss rates.
The main motivation to propose the effervescent model is to allow wide binary
companions to influence the morphology of the descendant PN.
For that, a wide binary companion should accrete mass, forms an accretion disks, and blows
jets (Soker 2001a).
The mass accretion rate and specific angular momentum of the accreted gas increase
with decreasing wind speed. Therefore, accretion from a slowly moving gas, with
velocities below the escape velocity from the AGB star, is a more efficient process
than accretion from a wind.

For that, and based on many observations, I suggest the existence of an extended zone
above the AGB star where parcels of gas do not reach the escape velocity (section 2).
The wind and bound gas exist simultaneously to distances of $\sim 100 \AU$.
This is termed the effervescent zone.
Under general assumption, I found that the density of the unbound material in that region
falls as $\sim r^{-5/2}$, not much faster than the wind density (section 3).
If the ejection rate of the unbound gas is similar to the mass loss rate of the wind,
then the companion basically accrete mass from the bound material in the effervescent zone
more than from the wind, up to orbital separation of $a \sim 100 \AU$.
This is sufficient to influence the shaping of the descendant PN (Akashi 2008).

The main result of the present paper is in strengthening previous claims (Soker 2001a)
that wide binary companions (with orbital separation of up to hundreds of AU) might
influence the shaping of the descendant PN.

The effervescent zone will contribute to lines emission at a very low speed. i.e.,
at the center of the line profiles obtained by the wind.
Decin et al. (2006) model the CO line emission from the red supergiant star \astrobj{VY CMa}.
Despite using three mass loss episodes, the wind still shows strong emission
near zero velocity, more than their model yields.
Similarly, Decin et al. (2007) model underestimates the line emission near zero velocity
in the case of the AGB star \astrobj{WX Psc}.
Decin et al. (2006) attribute the peak in emission of \astrobj{VY CMa} near zero velocity to
the acceleration zone of the wind at $r < 15 R_\ast \simeq 150 \AU$.
The present mass loss rate in their model of VY CMa is very high,
 $\sim 10^{-4} M_\odot \yr^{-1}$, favoring the existence of an effervescent zone.
I propose that this peak in emission near zero velocity, both in VY CMa and \astrobj{WX Psc},
can be attributed in part to an effervescent zone,
which overlaps with the acceleration zone of the wind.

I encourage observers and people conducting numerical simulations to further study the possible
existence of bound material simultaneously with the presence of a wind in the region
around AGB stars up to $\sim 100 \AU$ .

\acknowledgements
This research was supported by the Asher Space Research Institute in the Technion.

\end{document}